\documentclass[12pt]{article}
\usepackage{amsfonts}

\usepackage{amsmath}
\usepackage{graphicx}

\csname @addtoreset\endcsname{equation}{section}
\newcommand{\eq}{\begin{equation}}
\newcommand{\feq}{\end{equation}}
\newcommand{\eqn}{\begin{eqnarray}}
\newcommand{\feqn}{\end{eqnarray}}
\newcommand{\arr}{\begin{eqnarray*}}
\newcommand{\farr}{\end{eqnarray*}}

\input{tcilatex}

\begin{document}

\begin{titlepage}
\begin{flushright}
CAMS/01-04\\
\end{flushright}
\vspace{.3cm}
\begin{center}
\renewcommand{\thefootnote}{\fnsymbol{footnote}}
{\Large\bf
Curved Domain Walls of Five Dimensional Gauged Supergravity}
\vskip20mm
{\large\bf{A. H. Chamseddine \footnote{email: chams@aub.edu.lb} and
W.~A.~Sabra\footnote{email: ws00@aub.edu.lb}}}\\
\renewcommand{\thefootnote}{\arabic{footnote}}
\vskip2cm
{\it
Center for Advanced Mathematical Sciences (CAMS)
and\\
Physics Department, American University of Beirut, Lebanon.\\}
\end{center}
\vfill
\begin{center}
{\bf Abstract}
\end{center}
We study curved domain wall solutions for gauged supergravity theories
obtained by gauging some of the isometries of the manifold spanned by the
scalars of vector and hypermultiplets. We first consider the case  obtained by
compactifying M-theory on a Calabi-Yau threefold in the presence of G-fluxes.
It is found that supersymmetry allows for the construction of  domain wall
configurations with curved worldvolume and a cosmological constant. However it
turns out that the equations of motion, if one insists on the supersymmetric ansatz for the
scalars and warp factor, rule out solutions with a cosmological constant and
allows only for Ricci-flat worldvolumes.
Moreover, in the absence of flux, there are non-supersymmetric solutions
with worldvolumes given by Einstein manifolds. Our results are then  generalized to
all five dimensional gauged supergravity theories with vector and hypermultiplets.
\end{titlepage}

\section{\protect\bigskip Introduction}

Recently domain walls and black holes as solutions of five-dimensional
gauged supergravity have been a subject of intensive research. This to a
large extent has been motivated by the suggestion of Randall and Sundrum 
\cite{RS} that four dimensional Einstein gravity can be recovered provided
we live on a domain wall embedded in anti-de Sitter space. Ultimately the
aim is to embed such a model in string or M theory. The difficulty so far in
achieving this is due to the need of a specific model which gives rise to a
supersymmetric flow connecting two stable infra-red fixed points with the
same cosmological constant. In addition, explicit solutions of ungauged and
gauged supergravity theory provide the foundation for a microscopic
understanding of black hole physics and also play an important role in the
conjectured AdS/CFT correspondence\footnote[1]{%
For example, the AdS$_{5}\times S^{5}$ compactification of type IIB theory
gives $D=5$, $N=8$ gauged supergravity. The isometries of $S^{5}$ lead to a $%
SO(6)$ gauging, which in turn may be identified with the $SO(6)$ $R$%
-symmetry of the $D=4$, $N=4$ super-Yang-Mills theory on the boundary.} \cite
{ads}. Here, the anti-de Sitter geometry arises as the vacuum of gauged
supergravities in various dimensions. This correspondence may provide the
possibility to study the nonperturbative structure of the field theories
living on the boundary by means of classical supergravity solutions.
Supersymmetric black holes and strings,\cite{bcs1} as well as
non-supersymmetric generalizations \cite{bcs2}, have been constructed for
the $U(1)$ gauged $N=2$ supergravity\cite{gst2}. A specific model of these
theories, namely the STU model with three vectormultiplets constitutes a
consistent truncation of the $N=8$ theory.

These results motivate a further research in this field, in particular it is
of interest to find new solutions to general gauged supergravity theories
with non trivial vector and hypermultiplets scalars, such as black holes and
domain walls preserving some of the original supersymmetries. Also, one
would like to investigate generalizations of the gauged supergravity models
in order to find a model which may incorporate Randall-Sundrum scenario in a
supersymmetric setting. Domain walls and black hole solutions for gauged
theories with gauged isometries of the hypermultiplets have been discussed
very recently in \cite{domaink, gutsabra}. Curved domain wall solutions have
also been addressed in \cite{lust}.

In this paper we are mainly interested in the study of curved domain wall
solutions of gauged supergravity with vector and hypermultiplets. We
consider the models obtained from the compactification of M theory on
Calabi-Yau threefolds in the presence of background $G$-fluxes. In the
absence of fluxes, it is well known that the effective field theory obtained
is $N=2$ five-dimensional supergravity with hyper and vectormultiplets whose
number depends on the Betti numbers of the Calabi-Yau threefold. The scalar
fields of these multiplets parametrize a manifold $\mathcal{M=M}_{H}\mathcal{%
\otimes M}_{V}$. The scalars of the hypermultiplets live on a quaternionic
manifold $\mathcal{M}_{H},$ and those of the vectormultiplets live on $\ $a
very special real manifold $\mathcal{M}_{V}.$ In the presence of a
non-trivial $G$-flux, the axion becomes charged in the effective five
dimensional theory and one obtains a gauged supergravity model with a scalar
potential which depends on the volume of the Calabi-Yau as well as the
scalars of the vectormultiplets.

The main result of this work is that one may find supersymmetric field
configurations which preserve a fraction of supersymmetry but which do not
constitute solutions of the equations of motion. This should not come as a
surprise since it has already been established that in searching for
supersymmetric bosonic vacua, it is not enough to check for the existence of
parallel or Killing spinors. For instance, it was demonstrated in \cite
{aljose} that for Lorentzian spaces, Ricci-flatness is not an integrability
condition for the existence of parallel spinors. Moreover, from the analysis
of the equations of motion, it turns out that a domain wall solution with a
flat worldvolume, can be generalized to a solution where the worldvolume is
replaced with a Ricci-flat metric for the same values of the scalar fields
and warp factors. In the absence of flux, (i. e., ungauged five dimensional
supergravity) the equations of motion allow for non-supersymmetric solutions
with a positive cosmological constant. These results are then generalized to
any gauged supergravity model in five dimensions.

The paper is organized as follows. In the next section we study curved
domain wall solutions of M theory on a Calabi-Yau threefold in the presence
of background flux. These solutions correspond to M5-branes wrapped over
holomorphic curves in Calabi-Yau space. In section three we generalize our
results to all gauged supergravity theories with vector and hypermultiplets.
Finally we summarize and discuss our results.

\section{Curved Domain Walls}

In the following, we will study curved domain wall solutions of the
low-energy effective theory of M-theory on a Calabi-Yau threefold with
background flux. For details of the reduction, the reader is referred to 
\cite{lukas, bg}. This theory can also be obtained by gauging both the $U(1)$
subgroup of the $R-$symmetry and the axionic shift present in the universal
hypermultiplet\cite{ceresole, gutsabra}. We will consider the case where we
only keep the universal hypermultiplet. A hypermultiplet is present in any
Calabi-Yau compactification of M-theory and type II string theory. For
example, compactifying M-theory or type IIA string theory on a rigid
Calabi-Yau threefold ( i. e., $h_{2,1}=0$) leads to an $N=2$ theory with a
single hypermultiplet, the so-called universal hypermultiplet\footnote[2]{%
the term ``universal'' is slightly misleading for compactifications with $%
h_{2,1}>0$, see \cite{Aspinwall} section 4.4.3.} \cite{cfg}. Classically,
the scalar fields of the universal hypermultiplet parameterize the group
coset manifold $SU(2,1)/U(2)\cite{Ferrara,strominger}.$ Define the complex
coordinates\footnote[3]{%
In the above action, $\phi$ is associated with the volume of the Calabi-Yau, 
$a$ comes from the dual of the four-form of eleven dimensional supergravity, 
$F=dA_{3}$, and $C$ corresponds to the expectation values of $A_{3}.$}

\begin{equation}
S=e^{-2\phi}+ia+\chi_{1}^{2}+\chi_{2}^{2},\quad C=\chi_{1}+i\chi_{2}.
\end{equation}
The moduli space is K\"{a}hler with K\"{a}hler potential 
\begin{equation}
\mathcal{K}=\phi=-{\frac{1}{2}}\ln\Big({\frac{S+\bar{S}}{2}}-|C|^{2}\Big ). 
\label{eq:kaehlp}
\end{equation}

Using the coordinates $q^{X}=(S,\bar{S},C,\bar{C})$, the metric components
are 
\begin{equation}
g_{S\bar{S}}=\frac{1}{4}e^{4\phi},\text{ \ \ \ }g_{S\bar{C}}=-{{\frac{1}{2}C}%
}e^{4\phi},\text{ \ \ }g_{\bar{S}C}=-{{\frac{1}{2}}}\bar{C}e^{4\phi},\text{
\ \ \ }g_{C\bar{C}}=e^{2\phi}+C\bar{C}e^{4\phi}.
\end{equation}
The metric can be written as 
\begin{equation}
ds^{2}=u\otimes\bar{u}+v\otimes\bar{v},
\end{equation}
where we have introduced the vielbein forms by

\begin{align}
u & =e^{\mathcal{\phi}}dC,\quad  \notag \\
v & =e^{2\mathcal{\phi}}(\frac{dS}{2}-\bar{C}dC).
\end{align}
In addition to the hypermultiplet, the theory also contains the $N=2$
supergravity multiplet and $n_{V}$ vectormultiplets. The scalars $\phi^{x}$, 
$x=1,\ldots,n_{V}$ of the vectormultiplets parametrize a very special real
manifold $\mathcal{M}_{V}$ described by an $n_{V}$--dimensional cubic
hypersurface 
\begin{equation}
C_{IJK}h^{I}(\phi^{x})h^{J}(\phi^{x})h^{K}(\phi^{x})=1   \label{hypersurface}
\end{equation}
of an ambient space parametrized by $n_{V}+1$ coordinates $%
h^{I}=h^{I}(\phi^{x})$, where $C_{IJK}$ are the intersection numbers of the
Calabi-Yau threefold.

\bigskip For our domain wall solutions, the gauge fields are irrelevant and
therefore our Lagrangian is given by\footnote[4]{%
In this paper, the indices $A,B$ represent five-dimensional flat indices, $%
A=(a,5).$ Curved indices are represented by $M$ =$(\mu$, $z).$} 
\begin{equation}
\mathcal{L}=E\left( \frac{1}{2}R-\mathcal{\partial}_{M}\phi\mathcal{\partial 
}^{M}\phi-\frac{3}{4}G_{IJ}\mathcal{\partial}_{M}h^{I}\mathcal{\partial}%
^{M}h^{J}-\mathcal{V}(\phi,q)\right)   \label{axionaction}
\end{equation}
where 
\begin{equation}
\mathcal{V}={\frac{g^{2}}{8}}e^{4\phi}G^{IJ}\alpha_{I}\alpha_{J}
\end{equation}
is the scalar potential of the theory, $E=\sqrt{-\det g_{MN}}$ and $\alpha
_{I}$ are the flux vectors. We also have the following relations coming from
the underlying very special geometry\cite{dWvP}, 
\begin{align}
h^{I}h_{I}^{x} & =0,\qquad h_{x}^{I}h_{y}^{J}G_{IJ}=g_{xy},  \notag \\
G_{IJ} & =h_{I}h_{J}+h_{I}^{x}h_{J}^{y}g_{xy},  \notag \\
\partial_{x}h_{I} & =\sqrt{\frac{2}{3}}h_{Ix},\quad\partial_{x}h^{I}{}=-%
\sqrt{\frac{2}{3}}h_{x}^{I},   \label{qqq}
\end{align}
Note that only the field $\phi$ of the hypermultiplet is kept as a dynamical
variable. The supersymmetry transformations of the fermionic fields
(gravitini, gaugini and hyperini) in this model allow for the splitting of
the spinors components \cite{gutsabra} and for simplicity we will
concentrate on the equations for the spinor $\epsilon^{1}$ and drop the
spinor indices. Therefore the supersymmetry transformations we wish to study
in a bosonic background (and the absence of gauge fields) are

\begin{align}
\delta\psi_{M} & =\left( \partial_{M}+{{\frac{1}{4}}}\Omega_{M}^{AB}%
\Gamma_{AB}+{\frac{1}{4\sqrt{6}}}ge^{2\phi}\Gamma_{M}W\right) \varepsilon ,
\label{sone} \\
\delta\zeta & =-{\frac{i}{4}}e^{2\phi}\left(
\Gamma^{M}\partial_{M}e^{-2\phi}+{\frac{\sqrt{6}}{2}}gW\right) \varepsilon,
\label{stwo} \\
\delta\lambda_{x} & =\frac{i}{4}\left( 3\partial_{M}h_{I}\Gamma^{M}+\sqrt{%
\frac{3}{2}}ge^{2\phi}\alpha_{I}\right) \partial_{x}h^{I}\varepsilon. 
\label{sthree}
\end{align}

Here $W=\alpha_{I}h^{I}$ and $\Omega_{M}^{AB}$ are the spin connections. Our
purpose is to find supersymmetric domain wall solutions with curved
worldvolumes, this is achieved by solving the equations obtained by setting
the supersymmetry transformations of the fermi fields to zero. Solutions
with flat worldvolumes have been considered first in \cite{lukas}. The
metric of our curved brane can be put in the form

\begin{equation}
ds^{2}=e^{2U(z)}g_{\mu\nu}(x)dx^{\mu}dx^{\upsilon}+dz^{2}   \label{ansatz}
\end{equation}
and we assume that all the dynamical scalar fields of the theory depend only
on the fifth coordinate $z$. The non vanishing spin connections for our
metric are given by

\begin{align}
\Omega_{\mu ab}(x,z) & =\omega_{\mu ab}(x),  \notag \\
\Omega_{\mu a5}(x,z) & =U^{\prime}e^{U}e_{a\mu}(x).
\end{align}
Let us begin with the gravitino supersymmetry transformation (\ref{sone}).
Using our ansatz, this gives for the spatial components 
\begin{equation}
\delta\psi_{\mu}=\left( D_{\mu}+\frac{1}{2}U^{\prime}e^{U}\gamma_{\mu}%
\gamma_{5}+{\frac{1}{4\sqrt{6}}}ge^{2\phi}e^{U}\gamma_{\mu}W\right)
\varepsilon,   \label{gravv}
\end{equation}
where we have used 
\begin{equation}
\Gamma_{\mu}=e^{U}\gamma_{\mu},\ \ \Gamma_{z}=\gamma_{5},\ \ \ \ \ \ D_{\mu
}=\partial_{\mu}+\frac{1}{4}\omega_{\mu}^{ab}\gamma_{ab}.
\end{equation}
Here the prime denotes differentiation with respect to $z.$ From the fifth
component, we obtain

\begin{equation}
\delta\psi_{z}=\left( \partial_{z}+{\frac{1}{4\sqrt{6}}}ge^{2\phi}e^{U}%
\gamma_{5}W\right) \varepsilon.
\end{equation}
For the hyperino (\ref{stwo}), we obtain 
\begin{equation}
\delta\zeta=-{\frac{i}{4}}e^{2\phi}\left( \gamma_{5}\partial_{z}e^{-2\phi }+{%
\frac{\sqrt{6}}{2}}gW\right) \varepsilon.
\end{equation}
Assuming the following projection condition 
\begin{equation}
\gamma_{5}\varepsilon=-\varepsilon,   \label{fati}
\end{equation}
then the vanishing of the hyperino supersymmetry transformation implies the
following constraint 
\begin{equation}
gWe^{2\phi}=-\frac{4}{\sqrt{6}}\phi^{\prime}   \label{conone}
\end{equation}
which in turn implies for the vanishing of the gravitino transformations 
\begin{align}
\left( D_{\mu}-\frac{1}{6}e^{U}(\phi^{\prime}+3U^{\prime})\text{ }\gamma
_{\mu}\right) \varepsilon & =0,  \notag \\
\left( \partial_{z}+\frac{1}{6}\phi^{\prime}\right) \varepsilon & =0.
\end{align}
This implies that $\varepsilon=e^{-\frac{1}{6}\phi}\varepsilon(x)\varepsilon
_{0},$ where $\varepsilon_{0}$ is a constant spinor satisfying (\ref{fati})
and $\varepsilon(x)$ depends only on the worldvolume coordinates.
Integrability of the above equations implies that 
\begin{equation}
\frac{1}{6}e^{U}(\phi^{\prime}+3U^{\prime})=c,   \label{zin}
\end{equation}
where $c$ is a constant, we thus obtain

\begin{equation}
D_{\mu}\varepsilon(x)=\left( \partial_{\mu}+{{\frac{1}{4}}}%
\omega_{\mu}^{ab}(x)\gamma_{ab}-c\gamma_{\mu}\right) \varepsilon(x)=0.
\end{equation}
This is the Killing spinor equation for a purely gravitational background in
four dimensions. Therefore to obtain supersymmetric domain walls in five
dimensions, the four dimensional worldvolume must be a Lorentzian manifold
admitting Killing spinors. We shall come to this point later.

The vanishing of the gaugino supersymmetry variation yields

\begin{equation}
\left( \partial_{z}h_{I}-\frac{g}{\sqrt{6}}e^{2\phi}\alpha_{I}\right)
\partial_{x}h^{I}=0.
\end{equation}

This implies that the quantity in the bracket must be proportional to $h^{I}$
due to very special geometry. This will then imply that 
\begin{equation}
\left( \partial_{z}h_{I}-\frac{g}{\sqrt{6}}e^{2\phi}\alpha_{I}\right) =-%
\frac{g}{\sqrt{6}}e^{2\phi}Wh_{I}.
\end{equation}

This gives using (\ref{conone}), a simple differential equation 
\begin{equation}
\partial_{z}(e^{-\frac{2}{3}\phi}h_{I})=\sqrt{\frac{1}{6}}ge^{\frac{4}{3}%
\phi }\alpha_{I},   \label{conthree}
\end{equation}
which can be easily integrated if we perform the following change of
variable 
\begin{equation*}
\frac{dw}{dz}=e^{\frac{4}{3}\phi}
\end{equation*}
and the solution is given by 
\begin{equation}
e^{-\frac{2}{3}\phi}h_{I}=\frac{w}{\sqrt{6}}g\alpha_{I}+q_{I}
\end{equation}
where $q_{I}$ are constants. The above equation is normally referred to as
the attractor equation. Therefore the scalars of the theory can now be
determined in terms of algebraic equations. To fix the solution completely
(at least in terms of algebraic equations), we need to solve the
differential equation (\ref{zin}), 
\begin{equation*}
(\phi^{\prime}+3U^{\prime})=6ce^{-U}. 
\end{equation*}
This can be easily integrated using the solutions of the scalar fields to
give 
\begin{equation*}
e^{U(w)}=e^{-\frac{1}{3}\phi}\left( 2c\int e^{-\phi(w^{\prime})}dw^{\prime }+%
\text{const}\right) . 
\end{equation*}
To summarize, our domain wall configurations are given by 
\begin{align}
ds^{2} & =e^{2U(w)}g_{\mu\nu}(x)dx^{\mu}dx^{\nu}+e^{-\frac{4}{3}\phi}dw^{2},
\notag \\
Y_{I} & =h_{I}e^{-\frac{2}{3}\phi}=H_{I}=\frac{w}{\sqrt{6}}g\alpha_{I}+q_{I},
\notag \\
e^{-\phi} & =C_{IJK}Y^{I}Y^{J}Y^{K},\text{ \ \ \ }Y^{I}=e^{-\frac{1}{3}\phi
}h^{I}.   \label{summary}
\end{align}
This configuration is BPS if the four dimensional metric $g_{\mu\nu}(x)$
admits Killing spinors.

We now turn to the study of the equations of motion for our model and see
whether the above supersymmetric configuration is indeed a solution of the
theory. The Einstein equations obtained from the action (\ref{axionaction})
for our ansatz read

\begin{align}
R_{\mu\nu}^{(5)} & =R_{\mu\nu}^{(4)}-g_{\mu\nu}e^{2U}(U^{\prime\prime
}+4U^{\prime2})=\frac{2}{3}g_{\mu\nu}\mathcal{V}e^{2U},  \notag \\
R_{zz} & =2\phi^{\prime2}+\frac{3}{2}G_{IJ}\partial_{z}h^{I}%
\partial_{z}h^{J}+\frac{2}{3}\mathcal{V}=-4(U^{\prime\prime}+U^{\prime2}). 
\label{ein}
\end{align}
where ($R_{\mu\nu}^{(5)},$ $R_{zz})$ and $R_{\mu\nu}^{(4)}$ \ are the
Ricci-tensors of the domain wall and its four dimensional worldvolume
respectively. For the scalar field $\phi,$ related to the Calabi-Yau volume,
one gets

\begin{equation}
\frac{1}{E}\partial_{M}\left( Eg^{MN}\partial_{N}\phi\right) =\frac{1}{2}%
\frac{\partial\mathcal{V}}{\partial\phi}.   \label{seqm}
\end{equation}
To analyze Einstein equations of motion, we plug in our ansatz for the
scalar fields, this gives 
\begin{equation}
R_{\mu\nu}^{(4)}=3g_{\mu\nu}e^{2U}(U^{\prime2}-\frac{1}{9}\phi^{\prime2}).
\end{equation}
If we turn to the $\phi$ equation of motion, we obtain from (\ref{seqm}) 
\begin{equation}
\phi^{\prime\prime}+4\phi^{\prime}U^{\prime}=2\mathcal{V},
\end{equation}
where we have used $E=\sqrt{-\det g_{MN}}=e^{4U}\sqrt{-\det(g_{\mu\nu})}.$
Using (\ref{conone}), (\ref{conthree}) and (\ref{zin}) we conclude that $%
c=0, $ this yields 
\begin{align}
\phi^{\prime}+3U^{\prime} & =0,  \notag \\
R_{\mu\nu}^{(4)} & =0.
\end{align}
and hence the worldvolume metric of the domain wall solutions is Ricci-flat.
We conclude that supersymmetry allows for the possibility of supersymmetric
four dimensional worldvolume configurations with a negative cosmological
constant. However, these configurations satisfy the equations of motion only
if the cosmological constant is zero and the worldvolume is Ricci-flat.

A close inspection of the equations of motion reveals that if we consider
the supergravity cases where the potential $\mathcal{V}$ is zero, such as
ungauged supergravity models, then in the study of supersymmetric domain
walls, the scalar fields of the theory decouple and Einstein equations of
motion give,

\begin{align}
R_{\mu\nu}^{(5)} & =R_{\mu\nu}^{(4)}-g_{\mu\nu}e^{2U}(U^{\prime\prime
}+4U^{\prime2})=0,  \notag \\
R_{zz} & =-4(U^{\prime\prime}+U^{\prime2})=0.
\end{align}
In this case, the vanishing of the gravitino supersymmetry variation gives 
\begin{align}
\delta\psi_{\mu} & =\left( \partial_{\mu}+{{\frac{1}{4}}}\omega_{\mu}^{ab}%
\gamma_{ab}-\frac{1}{2}U^{\prime}e^{U}\gamma_{\mu}\right) \varepsilon =0, 
\notag \\
\delta\psi_{z} & =\partial_{z}\varepsilon=0
\end{align}
and integrability implies 
\begin{equation*}
U^{\prime}e^{U}=2c, 
\end{equation*}
which when plugged into the equations of motion gives 
\begin{equation*}
R_{\mu\nu}^{(4)}=12c^{2}g_{\mu\nu}. 
\end{equation*}
Therefore the equations of motion allow for Einstein spaces with positive
cosmological constant, clearly these solutions are not supersymmetric.
Curved supersymmetric solutions can be obtained in these cases if the
cosmological constant is zero.

\subsection{\protect\bigskip Supersymmetric curved solutions}

In this section, we return to the Killing spinor equation\footnote[5]{%
In the Mathematics literature, a Killing spinor equation refers to the case
with non-zero cosmological constant. In the absence of a cosmological
constant, our Killing spinors are known as parallel spinors.} that must be
satisfied in the four dimensional world

\begin{equation}
D_{\mu}\varepsilon(x)=\left( \partial_{\mu}+{{\frac{1}{4}}}%
\omega_{\mu}^{ab}(x)\gamma_{ab}\right) \varepsilon(x)=0.
\end{equation}
If one assumes static worldvolume, i. e, metrics admitting covariantly
constant time-like vector,

\begin{equation}
g_{\mu\nu}dx^{\mu}dx^{\upsilon}=-dt^{2}+ds_{3}^{2},
\end{equation}
then the only possibility for a supersymmetric solution is flat space \cite
{brecher}. On the other hand, considering metrics with a covariantly
constant light-like vector (parallel null-vector), then the most general $d$%
-dimensional Lorentzian metric in this case is known and was given many
years ago by Brinkmann \cite{brinkmann}. This is referred to as the
Brinkmann metric or the so-called $pp$-wave. In four dimensions, the general
form of this metric can be written in the following form 
\begin{equation}
ds^{2}=2dx^{+}dx^{-}+a(dx^{+})^{2}+b_{i}dx^{i}dx^{+}+g_{ij}dx^{i}dx^{j}.
\end{equation}

Here $i=1,..,d-2.$ The null vector is $\partial_{-}$, ($\partial_{-}a=%
\partial_{-}b_{i}=\partial_{-}g_{ij}=0).$ A subclass of these solutions with
covariantly constant spinors has been given in \cite{aljose}. The metric of
this subclass takes the form 
\begin{equation}
ds^{2}=2dx^{+}dx^{-}+a(dx^{+})^{2}+b%
\epsilon_{ij}x^{j}dx^{i}dx^{+}+dx^{i}dx^{i}   \label{hb}
\end{equation}

where $a=a(x^{+},x^{i})$ and $b=b(x^{+})$ are any two functions. An
mentioned in the introduction, it is important to note that the existence of
parallel spinors does not imply Ricci-flatness \cite{aljose}. In fact for
the above metric one finds that 
\begin{equation}
R_{++}=-\frac{1}{2}\Delta a+2b^{2}.
\end{equation}

where $\Delta$ =$\partial_{i}\partial_{i}$. In summary, the subclass (\ref
{hb}) of Brinkmann spaces have covariantly constant spinors without having
to satisfy Einstein equations of motion. Therefore it must be stressed that
in order to fix the supersymmetric bosonic solutions completely, one has to
solve the equations of motion and thus impose extra conditions. In this case
one must set $\ \Delta a=4b^{2}.$

\section{General case}

Gauged supergravity models are obtained when some global isometries of the
ungauged theory including $R$-symmetry are made local. Minimal $N=2$
supergravity theories, i. e, those with eight real supercharges, contain an
arbitrary number $n_{V\text{ }}$ of vectormultiplets and $n_{H}$
hypermultiplets (in this work tensor multiplets are ignored). The fermionic
fields of the theory are the two gravitini $\psi_{M}^{i}$ which are
symplectic Majorana spinors ($i=1,2$ are $SU_{R}(2)$ indices ), the gaugini $%
\lambda _{i}^{\hat{a}}$ \footnote[6]{%
the index $\hat{a}$ is the flat index of the tangent space group $SO(n_{V})$
of the scalar manifold $\mathcal{M}_{V}$} and the hyperini $\zeta^{\alpha}$ (%
$\alpha=1,...,2n_{H})$. The bosonic fields consist of the graviton, vector
bosons $A_{\mu}^{I}$ $(I=0,1,....,n_{V})$, the real scalar fields $\phi^{x} $
of the vectormultiplets and the scalars $q^{X}$ ($X=1,...,4n_{H})$ of the
hypermultiplet matter fields. The scalars of the theory parametrize a
manifold $\mathcal{M}$ which is the direct product of a very special and a
quaternionic manifold 
\begin{equation}
\mathcal{M}=\mathcal{M}_{V}\otimes\mathcal{M}_{H},
\end{equation}

The scalars $\phi^{x}$, $x=1,\ldots,n_{V}$, parametrize the target space $%
\mathcal{M}_{V}$. Note that for the quaternionic manifold, there are two
types of indices $\alpha$ and $i,$ corresponding to fundamental
representations of $USp(2n_{H})$ and $SU(2).$ The target manifold $\mathcal{M%
}_{V}$ of the scalar fields of the vectormultiplets is a very special
manifold described by an $n_{V}$--dimensional cubic hypersurface 
\begin{equation}
C_{IJK}h^{I}(\phi^{x})h^{J}(\phi^{x})h^{K}(\phi^{x})=1
\end{equation}
of an ambient space parametrized by $n_{V}+1$ coordinates $%
h^{I}=h^{I}(\phi^{x})$, where $C_{IJK}$ is a completely symmetric constant
tensor defining the Chern--Simons couplings of the vector fields. For more
details concerning the classification of the allowed homogeneous manifolds
we refer the reader to \cite{dWvP},\cite{gst1}.

The self-interacting scalars of the hypermultiplets in an $N=2$, $D=5$
theory live on a quaternionic K\"{a}hler manifold \cite{BW}, with a
quaternionic metric tensor which we denote by $g_{XY}(q)$.

The bosonic Lagrangian of the gauged theory for vanishing gauge fields is
given by 
\begin{equation}
\mathcal{L}=E\left( \frac{1}{2}R-\frac{1}{2}g_{XY}\partial_{M}q^{X}%
\partial^{M}q^{Y}-\frac{1}{2}g_{xy}\mathcal{\partial}_{M}\phi^{x}\mathcal{%
\partial}^{M}\phi^{y}-\mathcal{V}(\phi,q)\right) ,
\end{equation}

where the scalar potential is given by

\begin{equation}
\mathcal{V}=-g^{2}\left[ 2P_{ij}P^{ij}-P_{ij}^{\hat{a}}P^{\hat{a}\,ij}\right]
+2g^{2}\mathcal{N}_{i\alpha}\mathcal{N}^{i\alpha}.   \label{pot}
\end{equation}

Here 
\begin{align}
P_{ij} & \equiv h^{I}P_{I\,ij},  \notag \\
P_{ij}^{\hat{a}} & \equiv h^{\hat{a}I}P_{I\,ij},  \notag \\
\mathcal{N}^{i\alpha} & \equiv\frac{\sqrt{6}}{4}h^{I}K_{I}^{X}f_{X}^{\alpha
i},
\end{align}
Here $K_{I}^{X},P_{I\,}$ are the Killing vectors and prepotentials
respectively. The vielbeins $f_{i\alpha}^{X}$ obey the following relation $%
g_{XY}\,f_{i\alpha}^{X}\,f_{j\beta}^{Y}=\epsilon_{ij}\,C_{\alpha\beta}$,
where $\epsilon_{ij}$ and $C_{\alpha\beta}$ are the $SU(2)$ and $USp(2n_{H})$
invariant tensors respectively. For details of the gauging and notations we
refer the reader to \cite{ceresole}.

The bosonic part of the supersymmetry transformations of the fermi fields in
the gauged theory, after dropping the gauge fields contribution, are given
by 
\begin{align}
\delta_{\varepsilon}\psi_{Mi} & =\mathcal{D}_{M}\varepsilon_{i}+\frac {i}{%
\sqrt{6}}g\,\Gamma_{M}\varepsilon^{j}P_{ij},  \notag \\
\delta_{\varepsilon}\lambda_{i}^{\hat{a}} & =-\frac{i}{2}f_{x}^{\hat{a}%
}\Gamma^{M}\varepsilon_{i}\,\partial_{M}\phi^{x}+g\varepsilon^{j}P_{ij}^{%
\hat{a}},  \notag \\
\delta_{\varepsilon}\zeta^{\alpha} & =-\frac{i}{2}f_{iX}^{\alpha}\Gamma
^{M}\varepsilon^{i}\mathcal{\partial}_{M}q^{X}+g\varepsilon^{i}\mathcal{N}%
_{i}^{\alpha}.   \label{variation}
\end{align}
A general form for scalar potentials which guarantees stability \cite
{boucher} is given by 
\begin{equation}
\mathcal{V=}g^{2}(-6\mathcal{W}^{2}+\frac{9}{2}g^{\Lambda\Sigma}\partial_{%
\Lambda}\mathcal{W}\partial_{\Sigma}\mathcal{W})   \label{pottwo}
\end{equation}
where $\Lambda,\Sigma$ run over all the scalars of the theory. The
transition from (\ref{pot}) to (\ref{pottwo}) can be achieved by writing 
\footnote[7]{%
here $\mathcal{W}$ is the `superpotential' and $Q^{r}$ are $SU(2)$ phases}

\begin{equation}
h^{I}P_{I}^{r}=\sqrt{\frac{3}{2}}\mathcal{W}Q^{r},\text{ \ \ }Q^{r}Q^{r}=1
\end{equation}
and imposing the condition $\partial_{x}Q^{r\text{ }}=0$. This condition is
in general satisfied only on a submanifold of the total scalar manifold and
is also required for the existence of flat BPS domain wall solutions of the
theory \cite{domaink,proy}

Supersymmetric domain wall solutions with flat worldvolume of the form (\ref
{ansatz}) have been recently discussed in \cite{domaink}. The Killing
spinors of these solutions satisfy the projection condition 
\begin{equation}
\gamma_{5}\varepsilon_{i}=\sigma_{ij}^{r}Q^{r}\varepsilon^{j}.   \label{proj}
\end{equation}
It was found that for the flat BPS domain wall, the warp factor and the
scalar fields of the theory satisfy 
\begin{align}
\phi^{^{\prime}\Lambda} & =-3gg^{\Lambda\Sigma}\partial_{\Sigma}\mathcal{W},%
\text{ \ \ \ \ \ \ \ \ }\phi^{\Lambda}=(\phi^{x},q^{X}),  \label{scalars} \\
U^{\prime} & =g\mathcal{W}.   \label{metric}
\end{align}
We now generalize the results of the previous sections to all gauged
supergravity models with vector and hypermultiplets. It can be easily seen
that supersymmetry transformations allow for a negative cosmological
constant if one modifies (\ref{metric}) to take the form 
\begin{equation}
U^{\prime}=2ce^{-U}+g\mathcal{W}.   \label{modifiedmetric}
\end{equation}
The Einstein equations of motion give for the worldvolume Ricci-tensor the
following equation 
\begin{equation}
R_{\mu\nu}^{(4)}=g_{\mu\nu}e^{2U}(4U^{\prime2}+U^{\prime\prime}+\frac{2}{3}%
\mathcal{V)},   \label{general ricci}
\end{equation}
One also obtains for the Ricci-tensor $zz$-component , 
\begin{equation}
R_{zz}=g_{\Lambda\Sigma}\partial_{z}\phi^{\Lambda}\partial_{z}\phi^{\Sigma }+%
\frac{2}{3}\mathcal{V}=-4(U^{\prime\prime}+U^{\prime2})
\end{equation}
Using (\ref{scalars}) and (\ref{pottwo}) we then obtain 
\begin{equation}
(U^{\prime\prime}+U^{\prime2})=-3g^{2}g^{\Lambda\Sigma}\partial_{\Lambda }%
\mathcal{W}\partial_{\Sigma}\mathcal{W}+g^{2}\mathcal{W}^{2} 
\label{helpone}
\end{equation}
The relations (\ref{helpone}) and (\ref{pottwo}), when substituted in (\ref
{general ricci}) give 
\begin{equation}
R_{\mu\nu}^{(4)}=3g_{\mu\nu}e^{2U}(U^{\prime2}-g^{2}\mathcal{W}^{2}\mathcal{)%
}.
\end{equation}
Using (\ref{modifiedmetric}), we obtain that 
\begin{equation}
U^{\prime\prime}+U^{\prime2}=-3g^{2}g^{\Lambda\Sigma}\partial_{\Lambda }%
\mathcal{W}\partial_{\Sigma}\mathcal{W}+g^{2}\mathcal{W}^{2}+2cg\mathcal{W}
\end{equation}
Comparing with (\ref{helpone}), we deduce for non-vanishing $W$ that $c=0$
and therefore 
\begin{equation}
R_{\mu\nu}^{(4)}=0.
\end{equation}

One can easily check that the conditions (\ref{scalars}) and (\ref{metric})
solve the scalar fields equations of motion.

\section{\protect\bigskip Discussion}

We found general domain wall solutions of five dimensional gauged
supergravity theories coupled to vector and hypermultiplets. The examples
obtained from the compactification of M-theory on a Calabi-Yau threefold
were first examined and it was found that Killing spinor equations are
satisfied for configurations with anti-de Sitter worldvolumes. However, it
turns out that the equations of motion imply Ricci-flatness for the
worldvolume metric and thus exclude the case of non-zero cosmological
constant. Therefore, contrary to the common belief, supersymmetric solutions
are not necessarily solutions of the theory. Similar observation were
previously made in the literature. For example in \cite{aljose}, it was
shown that purely gravitational backgrounds can have parallel spinors
without having to satisfy Einstein equations of motion, i. e., not
Ricci-flat. In spaces with Lorentzian signature, having parallel spinors
does not imply Ricci flatness. Therefore extra conditions have to be imposed
in order for the space to be also a solution of Einstein gravity. Moreover,
it is known in the study of black holes (see \cite{bls}) that solving the
Killing spinors equations does not fix the bosonic supersymmetric solution
completely. There, one has to fix the solution in terms of harmonic
functions by solving for the gauge field equations of motion. Clearly,
harmonic functions can not be obtained from the first order differential
equations imposed by supersymmetry.

BPS curved domain wall solutions are obtained by searching for worldvolume
metrics with parallel spinors. The classification of space-times admitting
parallel spinors is a holonomy problem. In our case, i.e., for four
dimensional worldvolumes, one looks for subgroups of $Spin(3,1)=SL(2,C)$
which fix a spinor and then identify the Lorentzian manifolds which admit
these subgroups as their holonomy groups. This implies that the holonomy
group of the supersymmetric four-dimensional worldvolume is $\mathbb{R}^{2}$%
. Metrics with such a holonomy group are known and are given by the
supersymmetric $pp$-wave \cite{aljose}. Moreover, one can construct
non-supersymmetric curved domain wall solutions by having a worldvolume
metric which does not admit parallel spinors such as Schwarzschild black
hole. Non-supersymmetric solutions with Einstein worldvolumes with a
positive cosmological constant are possible in the cases of the ungauged
five dimensional supergravity models. It will be interesting to investigate
domain wall solutions with worldvolumes given by solutions with non-trivial
gauge fields \cite{branes}.

\textbf{Acknowledgments}. W. A. S thanks J. M. Figueroa-O'Farrill for many
useful discussions. \vfill\eject

\end{document}